\documentstyle[prb,aps,epsfig,floats,goodfloat]{revtex}

\newcommand{\be}{\begin{equation}}
\newcommand{\ee}{\end{equation}}
\newcommand{\beqn}{\begin{eqnarray}}
\newcommand{\eeqn}{\end{eqnarray}}
\newcommand{\nsw}{N_{\mathrm{sweep}}}
\newcommand{\nsa}{N_{\mathrm{samp}}}

\newcommand{\ql}{q_{l}}
\newcommand{\absq}{\left|q\right|}
\newcommand{\ml}{\mu_{l}}

\begin{document}
\draft

\twocolumn[
\hsize\textwidth\columnwidth\hsize\csname@twocolumnfalse%
\endcsname
\title{Nature of the spin-glass state in the three-dimensional gauge glass}
\author{Helmut G. Katzgraber and A. P. Young}
\address{Department of Physics, University of California, Santa Cruz, CA 95064}
\date{\today}
\maketitle
\begin{abstract}
We present results from simulations of the gauge glass model in
three dimensions using the parallel tempering Monte Carlo technique.
Critical fluctuations should not affect the data since we equilibrate down 
to low temperatures, for moderate sizes. Our results
are qualitatively consistent with earlier work on the three- and four-dimensional 
Edwards-Anderson Ising spin-glass. We find that large scale
excitations cost only a finite amount of energy in the thermodynamic limit, 
and that those excitations have a surface whose fractal dimension is less 
than the space dimension, consistent with a scenario proposed by Krzakala and
Martin, and Palassini and Young.
\end{abstract}
]

\section{Introduction}
\label{introduction}

The ground state structure of spin-glasses is poorly understood.
While there has been considerable work on Ising-type spin-glass
systems\cite{krzakala:00,palassini:00a,marinari:00,katzgraber:01},
models with a vector order parameter symmetry have not yet been analyzed 
in much detail. Here we study the 
three-dimensional gauge glass since it is a simple model with a vector order
parameter in which a finite-temperature spin-glass
transition\cite{olson:00} has been well established.

There are two theories describing the spin-glass phase: the ``droplet picture''
(DP) by Fisher and Huse\cite{fisher:87} and the replica symmetry breaking
picture (RSB) by Parisi\cite{parisi:79,mezard:87}. According to the droplet
picture, a
cluster of spins of size $L$ costs an energy proportional to $L^{\theta}$ 
where $\theta$ is positive. This implies that, in the thermodynamic limit,
excitations which flip a finite cluster of spins cost an infinite energy. In
addition, these excitations have a fractal surface with a fractal dimension
$d_s$ that is smaller than the space dimension $d$. By contrast, RSB follows 
the exact solution of the Sherrington-Kirkpatrick model in predicting that 
there are excitations which turn over a finite fraction of the
spins and which cost a finite amount of 
energy in the thermodynamic limit. The surface of these excitations is space 
filling\cite{marinari:00}, i.e.~$d_s = d$. Another difference
between these models can be quantified by looking at the distribution of the
order parameter\cite{marinari:00,reger:90,marinari:98a,zuliani:99} $P(q)$. 
In the droplet picture,  according to the standard
interpretation\cite{Bray86,Moore98}, $P(q)$ is trivial, i.e.~there
are only two peaks at $\pm q_{\rm EA}$in the thermodynamic limit ($q_{\rm EA}$ 
the Edwards-Anderson order parameter). For finite systems of linear size $L$, 
there is a tail with weight $\sim L^{-\theta}$ down to $q = 0$.
On the contrary, RSB predicts also a tail with a finite weight down to $q = 0$
independent of system size.

Recently, there have been results by Krzakala and Martin\cite{krzakala:00},
as well 
as Palassini and Young\cite{palassini:00a} (referred to as KMPY) for
Ising-type systems that find an intermediate picture: while large scale
excitations cost only a finite amount of energy in the thermodynamic limit,
their surface is fractal with $d_s < d$. In this scenario, it is necessary to
introduce two exponents, $\theta$ and $\theta'$,
to describe the system size dependence of the
excitation energy, where $L^\theta$ is the typical energy of an
excitation of size $L$ induced by a change in boundary
conditions, and $\theta^\prime$ describes the size dependence 
of the energy of clusters thermally excited at fixed boundary conditions.

In this paper we test which of the above predictions
apply to the gauge glass by performing Monte Carlo
simulations down to low temperatures
for a modest range of sizes using the parallel tempering Monte
Carlo method\cite{hukushima:96,marinari:98b}, as previously done in
Ref.~\onlinecite{katzgraber:01} for the three- and four-dimensional
Edwards-Anderson Ising spin-glass. We find that $P(0)$ does not decrease 
with increasing system size and, from data for $\ql$, we deduce that $d_s < d$, 
consistent with the existence of KMPY excitations. 

The layout of the paper is as follows: In Sec.~\ref{model-observables}
we describe the model as well as the observables measured while in
Sec.~\ref{equilibration} we discuss our equilibration tests for the parallel
tempering Monte Carlo method. Our results are discussed 
in Sec.~\ref{results}. In Sec.~\ref{conclusions} we summarize
our conclusions and present some ideas for future work. 

\section{Model and Observables}
\label{model-observables}

The Hamiltonian of the gauge glass is given by
\begin{equation}
{\cal H} = -J \sum_{\langle i, j\rangle} \cos(\phi_i - \phi_j - A_{ij}),
\label{hamiltonian}
\end{equation}
where the sum ranges over nearest neighbors on a square lattice in three
dimensions of size $N = L \times L \times L$ and $\phi_i$ represent
the angles of the XY spins. Periodic boundary conditions are applied.
$J$ is a positive ferromagnetic coupling between nearest neighbor spins
and $A_{ij}$ represents the line integral of the vector potential directed
from site $i$ to site $j$,
\begin{equation} 
A_{ij} = \frac{2\pi}{\Phi_0}\int_{{\bf r}_i}^{{\bf r}_j}{\bf A}\cdot
d{\bf l} ,
\label{gaugefields}
\end{equation}
and $\Phi_0 = h c/(2 e)$ is the flux quantum. The $A_{ij}$ are quenched random
variables uniformly distributed between $[0,2\pi]$. In this work we set
$J = 1$. Note that on average the gauge glass is isotropic even though there
are local quenched fluxes. 

This model is often used to describe disordered high-$T_c$
superconductors\cite{blatter:94} in a magnetic field since, even though it
lacks screening, it has the right order parameter symmetry.

The order parameter of the gauge glass is traditionally defined as
\begin{equation}
Q = {1 \over N} \sum_{i=1}^N
[ \langle e^{i \phi_i} \rangle' \langle e^{-i \phi_i}
\rangle']_{\rm av},
\label{eq:Q}
\end{equation}
where we indicate thermal averages by angular brackets,
$\langle \ldots \rangle$, and averages of over the disorder by rectangular
brackets, $[ \ldots ]_{\rm av}$.
If we average over all possible global
rotations of the spins then
the thermal averages in Eq.~(\ref{eq:Q}) vanish,
so we need to work in an ensemble where global rotations of are not permitted
(e.g. by applying a small symmetry breaking field along the $\phi = 0$ direction) 
for Eq.~(\ref{eq:Q}) to be
sensible. This is indicated by the prime on the thermal average.
In the simulation we evaluate the product of the two thermal
averages by simulating two copies (replicas) of the system with the same
quenched disorder, and so  Eq.~(\ref{eq:Q}) becomes
\begin{equation}
Q = [\langle \,q \,\rangle']_{\rm av} ,
\label{eq:Qq'}
\end{equation}
where the microscopic spin overlap
$q$ is defined by
\begin{equation}
q = \frac{1}{N}\sum_{j = 1}^{N}\exp\left(i (\phi^\alpha_i -
\phi^\beta_i)\right) ,
\label{overlap}
\end{equation}
where $\alpha$ and $\beta$ refer to the two replicas.
In the constrained ensemble, if we write $q\, (\equiv q_x +i q_y)$ in terms of
its real and imaginary parts, then $\langle q_y \rangle' = 0 $ and
$\langle q \rangle' =\langle q_x \rangle' $. 

In practice it is inconvenient to constrain the ensemble by applying a symmetry
breaking field, so instead we perform a rotation of one replica relative to
the other in order to maximize $q_x$ (and simultaneously set $q_y$ to zero).
It is easy to see that $q_x$ in the rotated frame is just $|q|$ (which is
invariant under rotations). We therefore take the spin order parameter to be
$|q|$ and its expectation value to be
\begin{equation}
Q = [\langle\, |q| \,\rangle] ,
\label{eq:Q|q|}
\end{equation}
(no prime now indicating an unconstrained thermal average) which is to be
compared with Eq.~(\ref{eq:Qq'}). In the unconstrained ensemble, the
probability to get particular values for $q_x, q_y$ only depends on $|q|$.

In addition, we will also study the link overlap $\ql$, defined by
\begin{equation}
\ql = \frac{1}{N_b}\sum_{\langle i,j \rangle}
\cos\left(
(\phi_i^\alpha - \phi_i^\beta) -
(\phi_j^\alpha - \phi_j^\beta)
\right) ,
\label{linkoverlap}
\end{equation}
where $N_b = N d$ is the number of bonds ($d = 3$ is the space dimension).
The sum ranges over all nearest-neighbor pairs of spins.
Note that while a change in $q$ induced by flipping a cluster of spins is
proportional to the {\it volume} of the cluster, $\ql$ changes by an
amount promotional to the {\it surface} of the cluster.

The weight in $P(\absq) $ for small $|q|$ varies as $L^{-\theta^\prime}$, 
where $\theta^\prime$ was introduced in section \ref{introduction}. 
In addition we expect the variance of the link overlap to 
fit to a form ${\rm Var}(\ql) \sim L^{-\mu_l}$ where, as shown in 
Ref.~\onlinecite{katzgraber:01}, $\mu_l = \theta^\prime +2(d - d_s)$.

\section{Equilibration}
\label{equilibration}

\begin{figure}
\centerline{\epsfxsize=\columnwidth \epsfbox{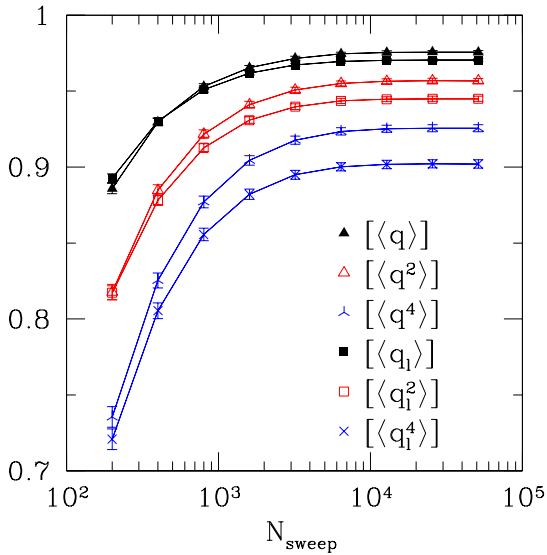}}
\caption{
Moments of the overlap and link overlap, defined by Eqs.~(\ref{overlap}) and
(\ref{linkoverlap}) respectively, as a function of
Monte Carlo sweeps $\nsw$, that each of the replicas perform, averaged
over the last half of the sweeps. Note that the moments seem
to equilibrate roughly at the same time and appear to be independent
of the number of sweeps. The data shown is for $L = 4$ and $T = 0.050$,
the lowest temperature studied.
}
\label{equil}
\end{figure}

For the simulations we use the parallel tempering Monte Carlo
method\cite{hukushima:96,marinari:98b} as it allows us to study
larger systems at lower temperatures.
In this technique, one simulates several identical replicas of the system
at different temperatures, and, in addition to the usual local moves, one
performs global moves in which the temperatures of two replicas (with
adjacent temperatures) are exchanged.
This method does not allow us
to use the equilibration test first introduced
by Bhatt and Young\cite{bhatt:85} 
because the system temperature does not stay constant throughout 
the simulation. The equilibration test for short range spin-glasses simulated
with parallel tempering Monte Carlo introduced
in Ref.~\onlinecite{katzgraber:01} 
by Katzgraber et.~al does not work here either because the disorder is not 
Gaussian. To ensure that the system is equilibrated, we therefore
require that different moments
of $q$ and $\ql$ are independent of the number of Monte Carlo steps $\nsw$.
Figure \ref{equil} shows data for several moments of $q$ and $\ql$ as a 
function of Monte Carlo Steps. One can clearly see that the different 
moments saturate at the {\em same} equilibration time.
Here we show  data for an intermediate size ($L=4$)
since we can better illustrate
the procedure by calculating longer equilibration times. 
We also require the acceptance ratios of the moves which interchange
temperatures 
to be at least $0.3$ or higher and roughly 
constant as a function of temperature.

\begin{figure}
\centerline{\epsfxsize=7cm \epsfbox{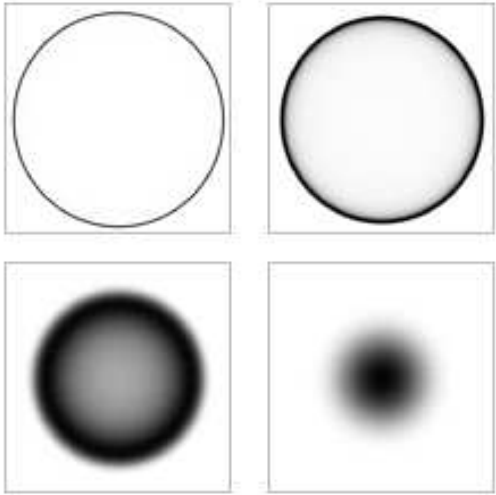}}
\caption{
Two-dimensional density plots of $P_{xy}(q_x,q_y)$ vs. $q_x$, $q_y$
for $L=4$. The
horizontal (vertical) axis represents $q_x$ ($q_y$). Dark corresponds to
a region
of high probability. Note that for 
different temperatures ranging from below to above $T_c$ the distributions 
are symmetric around $q = 0$. (Clockwise from top left: $T =$ 0.050, 0.166,
0.520 and 0.947).
}
\label{equil2d}
\end{figure}

As another test for equilibration,
we require the distribution of $q_x, q_y$, which we call $P_{xy}(q_x, q_y)$,
to be symmetric about the origin.
Figure \ref{equil2d} shows a density plot 
of $P_{xy}(q_x, q_y)$ for $L=4$ at different temperatures.
We see clearly that the distributions 
do not depend on the angle from the origin, as required.

In Table~\ref{simparams}, we show $\nsa$ (number of samples), $\nsw$
(total number of sweeps performed by each set of spins), and $N_T$
(number of temperature values), used in the simulations.
For each size, the largest temperature is 0.947 and the
lowest temperature is 0.050. This is to be compared with\cite{olson:00,reger:91}
$T_c \approx 0.45$. The set of temperatures is determined by requiring that 
the acceptance ratios for moves which exchange temperatures
are satisfactory for the largest size,
$L=8$, and for simplicity the same temperatures also are used for the smaller
sizes. We also want the distribution of $q$ at the highest temperature
to show a Gaussian shape centered at $q = 0$ to ensure that
all free-energy barriers have vanished
so the system will randomize quickly.
The lower-right image for $T = 0.947 \gg T_c$ in Figure \ref{equil2d}
shows this is the case.
For all system sizes, the acceptance ratios for global moves are
always greater than 0.3 for each pair of temperatures.

\begin{table}
\begin{center}
\begin{tabular}{lrrr}
$L$  &  $\nsa$  & $\nsw$ & $N_T$  \\ 
\hline
3 & $1.0 \times 10^4$ & $6.0 \times 10^3$ &   53 \\
4 & $1.0 \times 10^4$ & $2.0 \times 10^4$ &   53 \\
5 & $1.0 \times 10^4$ & $6.0 \times 10^4$ &   53 \\
6 & $5.0 \times 10^3$ & $2.0 \times 10^5$ &   53 \\
8 & $2.0 \times 10^3$ & $1.2 \times 10^6$ &   53
\end{tabular}
\end{center}
\caption{Parameters of the simulations. $\nsa$ is the
number of samples, i.e.~sets of gauge fields, $\nsw$ is the total number 
of sweeps simulated for each of the $2 N_T$ replicas for a single sample,
and $N_T$ is the number of temperatures used in the parallel tempering method.
}
\label{simparams}
\end{table}

Since the gauge glass has a vector order parameter symmetry, to speed up
the simulation we discretize the angles of the spins to $N_{\phi} = 512$.
This number is large enough to avoid any crossover effects to other models as 
discussed by Cieplak et.~al\cite{cieplak:92}.
To ensure a reasonable acceptance ratio for single-spin
Monte Carlo moves, we pick the proposed new angle for a spin
within an acceptance window about
the current angle, where
the size of the window is proportional to the temperature $T$. By tuning a 
numerical prefactor we ensure the acceptance ratios 
for these local moves are not smaller than 0.2 for each system size at the
lowest temperature simulated.

\begin{figure}
\centerline{\epsfxsize=\columnwidth \epsfbox{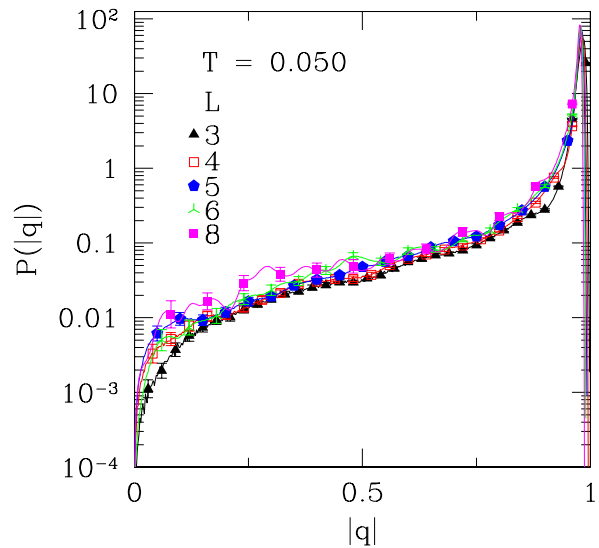}}
\caption{ Data for the overlap distribution
$P(\absq)$ at temperature $T = 0.050$ 
for different system sizes. Note the logarithmic vertical scale. 
In this and other similar figures in the paper, we only display a subset of
all the data points  while the lines connect all the data points in the set. 
Thus, the wiggles in the lines between neighboring symbols are meaningful.
}
\label{pq0.050}
\end{figure}

\section{Results}
\label{results}



\begin{figure}
\centerline{\epsfxsize=\columnwidth \epsfbox{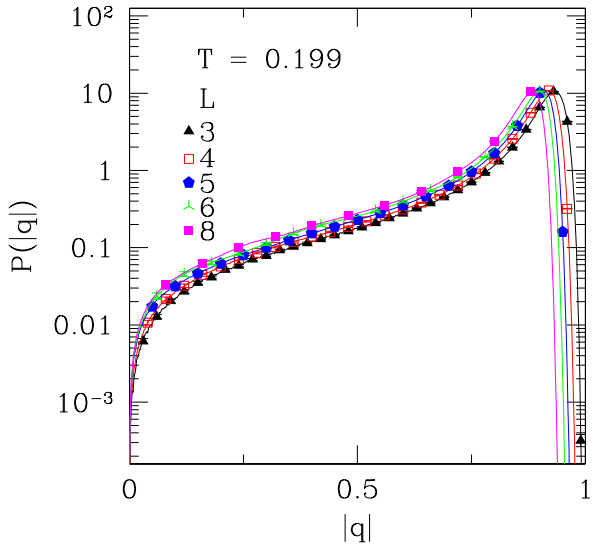}}
\caption{ Same as Figure \ref{pq0.050} but at temperature $T = 0.199$.}
\label{pq0.199}
\end{figure}

Figures \ref{pq0.050} and \ref{pq0.199} show data for $P(\absq)$ at $T =
0.050$ and $0.199$, well below\cite{olson:00} 
$T_c \approx 0.45$. 
There is a clear peak for large $\absq$ as well as a tail 
at small $\absq$. The weight in the tail does not decrease with increasing
$L$, as would be expected in the standard interpretations of the droplet
theory. If anything, the weight actually increases somewhat for larger sizes. 

Note that $P(\absq)$ decreases to zero, linearly, at very small $\absq$. This
is clearly a phase space factor, since the the two-dimensional
probability distribution $P_{xy}(q_x, q_y)$, plotted in Figure~\ref{equil2d},
will not diverge for a finite system, and $P(\absq) =
2 \pi \absq P_{xy}(q_x, q_y)$.
In order to have $P(\absq)$ tend to a constant for $\absq \to 0$, a
prediction of RSB, the region of $\absq$ over which $P(\absq)$ drops linearly
must tend to zero for $L \to \infty$, and also $P_{xy}(q_x, q_y)$ must diverge
as $1/\absq$ in this limit. There is some evidence, particularly from
Figure \ref{pq0.050}, that $P(\absq)$ stays flat down to smaller $\absq$ for
larger $L$, though the range of
sizes is too small to make a reliable extrapolation.

\begin{figure}
\centerline{\epsfxsize=\columnwidth \epsfbox{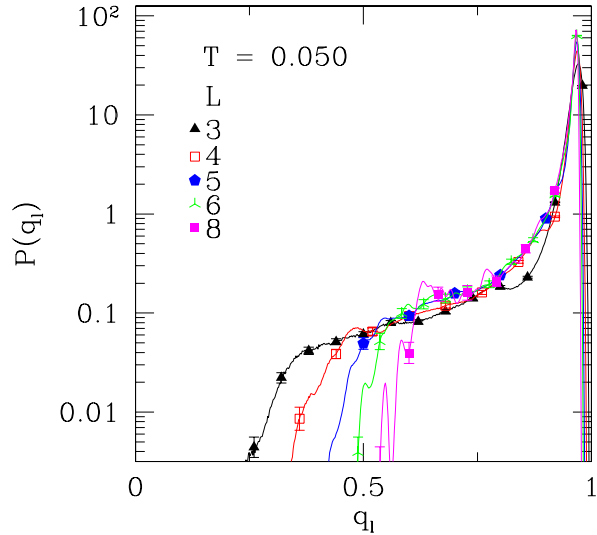}}
\caption{ The distribution of the link overlap at $T = 0.050$ for different
sizes. Note the logarithmic vertical scale.}
\label{pqb0.050}
\end{figure}
\begin{figure}
\centerline{\epsfxsize=\columnwidth \epsfbox{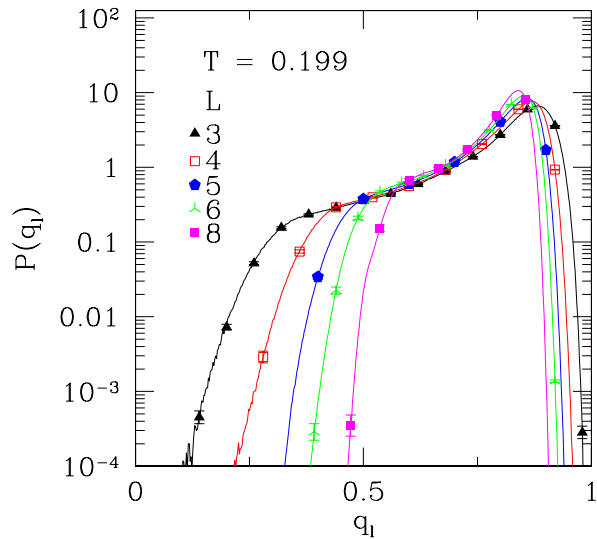}}
\caption{Same as Figure \ref{pqb0.050} but at temperature $T = 0.199$.}
\label{pqb0.199}
\end{figure}

Figures \ref{pqb0.050} and \ref{pqb0.199} show data for $P(\ql)$.
As with the distributions of $\absq$, there is a pronounced peak at large
$\ql$-values. The width of the distribution decreases with
increasing system size.

\begin{figure}
\centerline{\epsfxsize=\columnwidth \epsfbox{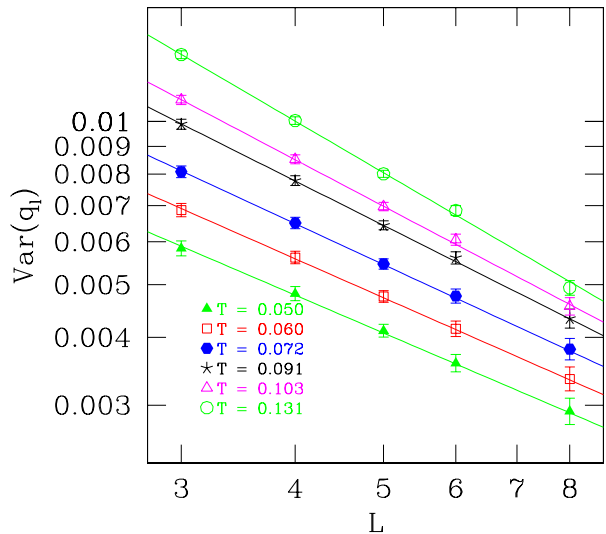}}
\caption{Log-log plot of the variance of $\ql$ as a function of system size 
$L$ at several temperatures.
}
\label{varqb}
\end{figure}

The variance of $P(\ql)$ is shown in Figure \ref{varqb} for several low
temperatures. The data is consistent with a power law decrease
where the (presumably effective) exponent
varies slightly with $T$. Extrapolating to $T = 0$ gives 
$\ml \equiv \theta^\prime + 2(d - d_s) = 0.501 \pm 0.04$.
Assuming $\theta^\prime \approx 0$ we find
\begin{equation}
d - d_s = 0.25 \pm 0.02,
\end{equation} 
implying that system-size excitations have a fractal surface in the
thermodynamic limit as predicted by the droplet picture.

\section{Conclusions}
\label{conclusions}

To conclude, Monte Carlo simulations of the three-dimensional gauge glass at
low temperatures show that the structure of the spin-glass state in this
particular model agrees qualitatively with previous results for Ising 
spin-glasses\cite{katzgraber:01} and is in agreement with the KMPY picture. 
From the non-trivial form of $P(q)$ for a  modest range of sizes we infer that
system-size excitations cost a finite amount of energy in the 
thermodynamic limit.
From the variance of the link overlap it appears that the
surface of these excitations is fractal with $d - d_s = 0.25 \pm 0.02$. 
Work is in progress on other vector spin-glass models to see if they have the
same features found here and for the Ising spin-glass.

These results, however, involve a large extrapolation to the 
thermodynamic limit. There may exist a crossover length at larger sizes to a 
different behavior such as the droplet theory or an RSB picture.

\acknowledgements

We would like to thank T.~Olson, for helpful discussions and correspondence.
This work was supported by the National Science Foundation under grant
DMR 0086287 and a Campus Laboratory Collaboration (CLC). 
The numerical calculations were made possible by use of the UCSC 
Physics graduate computing cluster funded by the Department of
Education Graduate Assistance in the Areas of National Need program.
We would also like to thank the University of New Mexico for access to their
Albuquerque High Performance Computing Center. This work utilized the 
UNM-Truchas Linux Clusters.


\begin{references}

\bibitem{krzakala:00}
    F.~Krzakala and O.~C.~Martin, Phys. Rev. Lett. {\bf 85}, 3013 (2000).

\bibitem{palassini:00a}
    M.~Palassini and A.~P.~Young, Phys. Rev. Lett. {\bf 85}, 3017 (2000).

\bibitem{marinari:00}
    E.~Marinari, G.~Parisi, F.~Ricci-Tersenghi, J.~Ruiz-Lorenzo and
    F.~Zuliani, J. Stat. Phys. {\bf 98}, 973 (2000).

\bibitem{katzgraber:01}
    H.~G.~Katzgraber, M.~Palassini and A.~P.~Young, Phys. Rev. B {\bf 63},
    184422, (2001).
 
\bibitem{olson:00}
    T.~Olson and A.~P.~Young, Phys. Rev. B {\bf 61}, 12467 (2000).

\bibitem{fisher:87}
    D.~S.~Fisher and D.~A.~Huse, J. Phys. A. {\bf 20} L997 (1987); D.~A.~Huse
    and D.~S.~Fisher, J. Phys. A. {\bf 20} L1005 (1987); D.~S.~Fisher and
    D.~A.~Huse, Phys. Rev. B {\bf 38} 386 (1988).

\bibitem{parisi:79}
    G.~Parisi, Phys. Rev. Lett. {\bf 43}, 1754 (1979); J. Phys. A {\bf 13},
    1101, 1887, L115 (1980; Phys. Rev. Lett. {\bf 50}, 1946 (1983).

\bibitem{mezard:87}
    M.~M\'ezard, G.~Parisi and M.~A.~Virasoro, {\em Spin glass Theory and
    Beyond} (World Scientific, Singapore, 1987).

\bibitem{reger:90}
    J.~D.~Reger, R.~N.~Bhatt and A.~P.~Young, Phys. Rev. Lett. {\bf 64}, 1859
    (1990).

\bibitem{marinari:98a}
    E.~Marinari, G.~Parisi, and J.~J.~Ruiz-Lorenzo, in {\em Spin glasses and
    Random Fields}, edited by A.~P.~Young (World Scientific, Singapore, 1998),
    and references therein.

\bibitem{zuliani:99}
    E. Marinari and F. Zuliani, J. Phys. A {\bf 32}, 7447 (1999).

\bibitem{Bray86} 
    A.~J. Bray and M.~A.~Moore,  in {\em Heidelberg Colloquium on Glassy Dynamics and
    Optimization}, edited by L. Van~Hemmen and I. Morgenstern
    (Springer-Verlag, Berlin, 1986), p.\ 121.

\bibitem{Moore98}
    M.~A. Moore, H. Bokil, and B. Drossel, Physical Review Letters {\bf 81},
    4252 (1998).

\bibitem{hukushima:96}
    K.~Hukushima and K.~Nemoto, J. Phys. Soc. Japan {\bf 65}, 1604
    (1996).

\bibitem{marinari:98b}
    E.~Marinari, {\em Advances in Computer Simulation}, edited by J. Kert\'esz
    and I.~Kondor (Springer-Verlag, Berlin 1998), p. 50, (cond-mat/9612010).

\bibitem{blatter:94}
    For a review of vortices in superconductors, see G.~Blatter,
    M.~V.~Feigel'man, V.~B.~Geshkenbein, A.~I.~Larkin and ~V.~M.~Vinokur,
    Rev. Mod. Phys. {\bf 66},
    1125 (1994).

\bibitem{bhatt:85}
    R.~N.~Bhatt and A.~P.~Young, Phys. Rev. Lett. {\bf 54}, 924 (1985);
    {\em ibid.}\/, Phys. Rev. B {\bf 37}, 5606 (1988).

\bibitem{reger:91}
    J.~D.~Reger, T.~A.~Tokuyasu, A.~P. Young and M.~P.~A. Fisher,
    Phys. Rev. B {\bf 44}, 7147 (1991).

\bibitem{cieplak:92}
    M.~Cieplak, J.~.R.~Banavar, M.~S.~Li and A.~Khurana, 
    Phys. Rev. B {\bf 45}, 786 (1992).

\end{references}
\end{document}